%% file: quant-ph9902068
\documentclass[11pt,titlepage,a4paper]{article}
\usepackage{amsfonts,amsbsy,amsopn}	

\usepackage{bozhomac}	
\usepackage{bozhlogo}

\title{\bfseries	\vspace*{-2.178902345in}
\vspace*{-3ex}
{
\begin{flushright}
	\textbf{\large LANL xxx E-print archive No. quant-ph/9902068}\\[2ex]
\end{flushright}
}
{\huge Fibre bundle formulation of 	\\[6pt]
nonrelativistic quantum mechanics	\\[2ex]
\hspace*{-1.123456ex}\Large
	V. Theory's interpretation, \\ summary and discussion
}
}

\vspace{3ex}

\author{
Bozhidar Z. Iliev
\thanks{Department Mathematical Modeling,
Institute for Nuclear Research and \mbox{Nuclear} Energy,
Bulgarian Academy of Sciences,
Boul. Tzarigradsko chauss\'ee~72, 1784 Sofia, Bulgaria}
\thanks{E-mail address: bozho@inrne.bas.bg}
\thanks{URL: http://www.inrne.bas.bg/mathmod/bozhome/}
}

\date{
\vspace{2.27ex}\ShortTitle{Bundle quantum mechanics: V}	\\[0.27ex]
\vspace{3.27ex}
	\begin{tabular}{r@{$\colon\to~$}l}
\vspace{0.09ex} Basic ideas	& March 1996		\\[0.09ex]
\vspace{0.09ex} Began		& May 19, 1996		\\[0.09ex]
\vspace{0.09ex} Ended		& July 12, 1996		\\[0.09ex]
\vspace{0.09ex} Revised		& December 1996 -- January 1997,\\[0.09ex]
\vspace{0.09ex} Revised		& April 1997, September 1998	\\[0.09ex]
\vspace{0.09ex} Last update	& November 16, 1998  	\\[0.09ex]
\vspace{0.09ex} Composing/Extracting part V
			& September 27/October 4, 1997	\\[0.09ex]
\vspace{0.09ex} Updating part V & February 20, 1999  	\\[0.09ex]
\vspace{0.27ex} Produced	& \fbox{\today}		\\[0.27ex]
	\end{tabular} \\[1.27ex]
	\begin{tabular}{r@{$\colon~$}l}
\vspace{0.27ex} LANL xxx archive server E-print No.& quant-ph/9902068
							\\[0.27ex]
	\end{tabular} \\[-0.27ex]
\vspace{3.27ex}{\Huge\BOZHO}	\\[3.27ex]
\vspace{0.27ex}\Subject{Quantum mechanics; Differential geometry} \\[2.27ex]
	\begin{tabular}{r@{\hspace{0.512em}}|@{\hspace{0.512em}}l}
\vspace{0.27ex}		  
&
\vspace{0.27ex}
	\end{tabular} \\[1.27ex]
\vspace{0.27ex}	\\[0.27ex]
}
\newcommand{\Hil}{\mathcal{F}}	
\newcommand{\HilB}{(\bHil,\proj,\base)}	
	\newcommand{\bHil}{\mathit{F}}	
	\newcommand{\proj}{\pi}		
	\newcommand{\base}{\mathit{M}}	

\newcommand{\Ham}{\mathcal{H}}	
\newcommand{\bHam}{\mathit{H}}	

\newcommand{\mHam}{\boldsymbol{\Ham}^\mathbf{m}}   
\newcommand{\mbHam}{\boldsymbol{\bHam}\!^\mathbf{m}}

\newcommand{\dyn}[1]{\pmb{\mathbb{#1}}}	
	\newcommand{\ope}[1]{\mathcal{#1}}		 
	\newcommand{\mope}[1]{\boldsymbol{\mathcal{#1}}} 
	\newcommand{\mor}[1]{\mathit{#1}}		 
	\newcommand{\mmor}[1]{\boldsymbol{\mathit{#1}}}	 

\newcommand{\ih}{\mathrm{i}\hbar}

\newcommand{\Rho}{\mathrm{P}} 

\begin{document}		


\renewcommand*{\thesection}{I.\arabic{section}}
\renewcommand*{\thefootnote}{\protect{I.\arabic{footnote}}}
\setcounter{page}{0}
\setcounter{footnote}{0}
\setcounter{section}{0}
\setcounter{equation}{0}
 	\include{bqm-1txt}
\renewcommand*{\thesection}{\arabic{section}}
\renewcommand*{\thefootnote}{\arabic{footnote}}
\setcounter{page}{0}
\setcounter{footnote}{0}
\setcounter{section}{0}
\setcounter{equation}{0}

\renewcommand*{\thesection}{II.\arabic{section}}
\renewcommand*{\thefootnote}{\protect{II.\arabic{footnote}}}
\setcounter{page}{0}
\setcounter{footnote}{0}
\setcounter{section}{0}
\setcounter{equation}{0}
 	\include{bqm-2txt}
\renewcommand*{\thesection}{\arabic{section}}
\renewcommand*{\thefootnote}{\arabic{footnote}}
\setcounter{page}{0}
\setcounter{footnote}{0}
\setcounter{section}{0}
\setcounter{equation}{0}

\renewcommand*{\thesection}{III.\arabic{section}}
\renewcommand*{\thefootnote}{\protect{III.\arabic{footnote}}}
\setcounter{page}{0}
\setcounter{footnote}{0}
\setcounter{section}{0}
\setcounter{equation}{0}
 	\include{bqm-3txt}
\renewcommand*{\thesection}{\arabic{section}}
\renewcommand*{\thefootnote}{\arabic{footnote}}
\setcounter{page}{0}
\setcounter{footnote}{0}
\setcounter{section}{0}
\setcounter{equation}{0}

\renewcommand*{\thesection}{IV.\arabic{section}}
\renewcommand*{\thefootnote}{\protect{IV.\arabic{footnote}}}
\setcounter{page}{0}
\setcounter{footnote}{0}
\setcounter{section}{0}
\setcounter{equation}{0}
 	\include{bqm-4txt}
\renewcommand*{\thesection}{\arabic{section}}
\renewcommand*{\thefootnote}{\arabic{footnote}}
\setcounter{page}{0}
\setcounter{footnote}{0}
\setcounter{section}{0}
\setcounter{equation}{0}


\renewcommand{\thefootnote}{\fnsymbol{footnote}}
\maketitle
\renewcommand{\thefootnote}{\arabic{footnote}}

\tableofcontents		


\pagestyle{myheadings}
\markright{\itshape\bfseries Bozhidar Z. Iliev:
	\upshape\sffamily\bfseries Bundle quantum mechanics.~V}

\begin{abstract}

We propose a new systematic fibre bundle formulation of nonrelativistic
quantum mechanics. The new form of the theory is equivalent to the usual one
but it is in harmony with the modern trends in theoretical physics and
potentially admits new generalizations in different directions. In it
a pure state of some quantum system is described by a state section (along
paths) of a (Hilbert) fibre bundle. It's evolution is determined through the
bundle (analogue of the) Schr\"odinger equation. Now the dynamical variables
and the density operator are described via bundle morphisms (along paths).
The mentioned quantities are connected by a number of relations derived in
this work.
%
%

	This is the fifth closing part of our investigation. We briefly
discuss the observer's r\^ole in the theory and different realizations of the
space-time model used as a base space in the bundle approach to quantum
mechanics. We point the exact conditions for the equivalence of Hilbert
bundle and Hilbert space formulations of the theory. A comparison table
between the both description of nonrelativistic quantum mechanics is
presented.  We discuss some principal moments of the Hilbert bundle
description and show that it is more general than the Hilbert space one.
Different directions for further research are pointed too.

\end{abstract}

\section {Introduction}
\label{introduction-V}
\setcounter{equation} {0}

	The present paper is the fifth closing part of our series of
works~\cite{bp-BQM-introduction+transport,bp-BQM-equations+observables,
			bp-BQM-pictures+integrals,bp-BQM-mixed_states+curvature}
devoted to the fibre bundle formulation of nonrelativistic quantum mechanics.
It is organized as follows.

	In Sect.~\ref{XII} is paid attention on the observer's r\^ole in the
theory and are considered and interpreted some modifications of the proposed
approach to quantum mechanics. Possible fields for further research are
sketched too.

	In Sect.~\ref{summary} are briefly summarized our
results and  it is presented a comparison table between the conventional,
Hilbert space, and the new, Hilbert bundle, formulations of
nonrelativistic quantum mechanics.

	In Sect.~\ref{discussion} are discussed certain aspects of the bundle
formulation of nonrelativistic quantum mechanics and are pointed some its
possible generalizations and applications.

\vspace{1.2ex}

	The notation of the present part of the series is identical with
the one of the preceding ones for which the reader is referred
to~\cite{bp-BQM-introduction+transport,bp-BQM-equations+observables,
			bp-BQM-pictures+integrals,bp-BQM-mixed_states+curvature}.

	The references to sections, equations, footnotes etc.
from the previous four parts of the series,
namely~\cite{bp-BQM-introduction+transport},
	\cite{bp-BQM-equations+observables},
	\cite{bp-BQM-pictures+integrals},
	and~\cite{bp-BQM-mixed_states+curvature},
are denoted by the corresponding sequential reference numbers in these parts
preceded by the Roman number of the part in which it appears and a dot as a
separator. For instance, Sect.~I.5 and (IV.2.11) mean respectively
section 5 of part~I, i.e.\  of~\cite{bp-BQM-introduction+transport}, and
equation~(2.11) (equation~11 in Sect.~2) of
part~IV, i.e.\  of~\cite{bp-BQM-mixed_states+curvature}.

\section{On observer's r\^ole and theory's interpretation}
\label{XII}

	The concept of an `observer' is more physical than mathematical one
as it is not very well mathematically rigorously defined; some times its
meaning is more intuitive than strict one. Generally an `\emph{observer}' is
a physical system whose state is assumed to be `completely' known and which
has a double r\^ole with respect to the other systems(s) which is (are) under
consideration. From one hand, it provides certain reference point, generally
a set of objects and their properties, with respect to which are determined
(all of) the quantities characterizing the investigated system(s) in some
problem. From the other hand, it is supposed the `observer' can perform
certain procedures, called `measurements' or `observations', by means of
which `he' finds (determines) the parameters, properties, quantities, etc.
describing the studied system(s). This second r\^ole of the observers is out
of the subject of the present investigation and will not be discussed here
(for some its aspects see, e.g.,~\cite{Messiah-QM,Prugovecki-QMinHS}).

	In this work the observers are supposed to be local and point-like,
i.e.\   they are material points that can perform measurements at the points at
which they are situated. They are moving (along paths) in some differentiable
manifold $\base$.%
\footnote{%
Mathematically the developed here theory is sensible also if $\base$ is
considered as a `more general' object than a manifold, but, at present, there
are not indications that such a theory can be physically important.%
}
Namely their trajectories (world lines in special or general relativity
interpretation (see below)) are the reference objects with respect to which
we study the behaviour of the quantum systems. The `observational' properties
of an observer are assumed fixed and such that: (i) Allow the observer to
determine the initial values of the quantities characterizing the state of
the studied system(s) at some instant of time $t_0$; (ii) Give certain
correspondence rules according to which to any dynamical variable $\dyn{A}$,
connected to the investigated system(s), is assigned some observable which is
a Hermitian operator (resp.\ morphism along paths), say $\ope{A}$ (resp.\
$\mor{A}$), in the Hilbert space (resp.\ bundle) description that has a
complete set of (maybe orthonormal) eigenvectors (resp.\ eigensections along
paths).

	All quantities in the present work are referred to an observer moving
along some path $\gamma\colon J\to\base$ parameterized with $t\in J$, where
$J\subseteq\mathbb{R}$  is a real interval. Our intention is to interpret
$t$ as a `time'. The possibility for such an interpretation is connected with
the specific choice of the manifold $\base$.

If we assume $\base$ to be the classical, coordinate, 3-dimensional
Euclidean space $\mathbb{E}^3$ of the classical and quantum mechanics, which
we have done more implicitly than explicitly throughout this investigation,
there exists a \emph{global time}
$t_\mathrm{g}\in(-\infty,+\infty)=\mathbb{R}$
(in the Newtonian sense). In this case the trajectories, such as $\gamma$,
of \emph{all} observers is natural and convenient to be parameterized
with this global time which, in fact, is done in the conventional, Hilbert
space, quantum mechanics. (One can also parameterize each observer's
trajectory with its own local time which is in one-to-one, usually $C^1$,
correspondence with $t_\mathrm{g}$, but this is an inessential
generalization as $t_\mathrm{g}$ itself is defined with some arbitrariness
(usually a $C^1$ map $\mathbb{R}\to\mathbb{R}$).)
Consequently the
assumptions $\base=\mathbb{E}^3$, $J=\mathbb{R}$ and $t=t_\mathrm{g}$ are
necessary and sufficient for the full equivalence of the Hilbert space and
Hilbert bundle descriptions of quantum mechanics. These assumptions lead to
one inessential mathematical complication. Since the observer's trajectory
$\gamma$ can has self-intersections, the sections along $\gamma$ and
morphisms along $\gamma$ can be multiple-valued at the points of
self-intersection of $\gamma$, if any. But this does not have some serious
consequences.

	Analogous to $\base=\mathbb{E}^3$ is the case when the Hilbert bundle
is identified with the system's configuration space. The only difference is
that $\gamma\colon J\to\base$ has to be interpreted as the trajectory of the
system in this space rather then the one of some observer. Practically the
same is the case when as $\base$ is taken the system's phase space but, since
this situation has some peculiarities, we shall comment on it in
Sect.~\ref{discussion}.

	A similar, but slightly different, is the situation when $\base$ is
taken to be the four dimensional Minkowski space-time $M^4$ of special
relativity.%
\footnote{%
A like construction, under the name `Schr\"odinger bundle', is introduced in
the paragraph containing equation~(4) of~\cite{Aldaya&Azcarraga-81}. This is
a Hilbert bundle over $M^4$ having as a (standard) fibre
$\Hil\times\Hil^\ast$ instead of the conventional Hilbert space $\Hil$ in our
case.%
}
Nevertheless that now every observer has its own local preferred
time, called proper or eigen-time, there is a global (e.g.\ coordinate) time
$t_\mathrm{g}$ which is in 1:1 correspondence with these local times. In this
case the observers are moving along paths, such as $\gamma$, which are their
\emph{world lines (paths or curves)} and usually are parameterized with the
global time $t_\mathrm{g}$. This is an important moment because the world
lines of the real objects observed until now can not have self-intersections.
Therefore, since the observers are supposed to be such, their world lines are
without self-intersections. This implies the absence of the complication
mentioned at the end of the previous paragraph, viz.\ now the sections along
$\gamma$ and morphisms along $\gamma$ are single-valued and, in fact, are
respectively sections and morphisms of the restriction
$\left.\HilB\right|_{\gamma(J)}$ of the Hilbert bundle $\HilB$ on the set
$\gamma(J)$. Otherwise the case $\base=M^4$ is identical with the one with
$\base=\mathbb{E}^3$. Therefore we can say that it represents the Hilbert
bundle description of the nonrelativistic quantum mechanics over the
special relativity space-time.  This point is worth-mentioning as at it meet
the relativistic and non-relativistic concepts whose unification leads to the
relativistic quantum theory which will be considered elsewhere.

	Another important possibility is $\base=V_4$ where  $V_4$ is the four
dimensional pseudo-Riemannian space-time of general relativity.%
\footnote{%
Almost the same is the case when $\base$ represents the space-time model of
other gravitational theories, like Einstein-Cartan and the metric-affine
ones, in which $\base$ is a curved (non-flat) manifold with respect to some
linear connection.%
}
The crucial point here is the generic non-existence of some global time, so the
\emph{world line} of any particular observer, say $\gamma$, with necessity
has to be parameterize with its (specifically local) \emph{proper time} $t$.
A consequence of this is that in the (conventional) Hilbert space description
the parameter (`time') $t$ also has to be considered as a local (proper) time
for the observer which describes the quantum system under consideration.
Hence the global sections and morphisms defined via~\eqref{4.3b}
and~\eqref{6.1} have no physical meaning now. As in special relativity case,
$\base=M^4$, now $\gamma$ cannot have self-intersections; so the sections and
morphisms along $\gamma$ are single-valued. What concerns other aspects of
the case $\base=V_4$, it is identical with the one for $\base=\mathbb{E}^3$.
Consequently, it represents the non-relativistic quantum mechanics over the
general relativity space-time. Here we see again a meeting of (general)
relativistic and nonrelativistic concepts whose unification will be
investigated elsewhere.

	Now we want to call attention to a particular `degenerate' case which
falls out of our general interpretation of $\gamma:J\to M$ as an observer's
world line (trajectory). Namely, it is possible to put $\base=J$, $J$ being a
real interval. For instance, as we pointed at the end of Sect.~\ref{V}, the
considerations of~\cite{Asorey&Carinena&Paramio} correspond to the choice
$\base=\mathbb{R}_+=\{t: t\in\mathbb{R},\ t\ge0\}$. Thus, if $\base=J$,
then $\gamma:J\to J$ and, as we tend to interpret $t\in J$ as time, it is
natural now to assume that $\gamma$ is one-to-one smooth ($C^1$) map.%
\footnote{%
The case considered in reference~\cite{Asorey&Carinena&Paramio} corresponds
to $\gamma=\id_{\mathbb{R}_+}$.%
}
If so, $\gamma(t)\in J$ is also a `presentation' of the time, but in
other (re)parametrization. In this case $\gamma$ is without
self\ndash intersections and, consequently, the sections and morphisms along
$\gamma$ are simply (single-valued) sections and morphisms of the bundle
$(\bHil,\proj,J)$ over the one\ndash dimensional base $J$. Therefore, in this
situation, we can say that the evolution of a quantum system is described via
linear transportation (of the state) sections of $(\bHil,\proj,J)$ along the
time, respectively the evolution of the observables is represented by linear
transportation of (the observable) morphisms of $(\bHil,\proj,J)$ along the
time. Let us note that now the connection with the observers is not
completely lost as the time does not exist by `itself' in the theory; it is
connected with (measured by) a concrete observer regardless of the fact that
it can be global of local (see above).

	Another degenerate case is when $\base$ consist of a single point,
$\base=\{x\}$. Then $\bHil=\proj^{-1}(x)=:\bHil_{x}$ and
$\gamma\colon J\to\{x\}$. Therefore $\gamma(t)\equiv x$ for every $t\in J$
and, if $J$ is not compressed into a single real number, $\gamma$
self-intersects at $x$ infinitely many times. Also we have
 $l_{\gamma(t)}=l_{\gamma(s)}=l_{x}=:l$, $s,t\in J$ with
$l\colon\bHil\to\Hil$ being an isomorphism. So, in this way (see the paragraph
after the one containing equation~\eref{4.3a}), we obtain an isomorphic image
in $\bHil$ of the quantum mechanics in $\Hil$. Evidently, the conventional
quantum mechanics is recovered by the choices $\Hil=\bHil$ and
$l=\id_{\bHil}$. Of course, now we can not interpreted $\gamma$ as observer's
trajectory or world line but the interpretation of $t$ as a `time' can be
preserved.

	If the quantum system under consideration has a classical analogue,
then the manifold $\base$ can be identified with the system's configuration
or phase space. In this case the path $\gamma\colon J\to\base$ can be taken
to be the trajectory of the system's classical analogue in the corresponding
space. Thus we obtain an interesting situation: the (bundle) quantum evolution
is described with respect to (is referred to) the corresponding classical
evolution of the same system.

	One can also take $\base$ to be the configuration or phase space of
some observer. Then $\gamma$ can naturally be defined as the observer's
trajectory in the corresponding space.

	At this point we want to say a few words on the possibility to
identify the Hilbert bundle's base $\base$ with the phase space of certain
system and to make some comments on~\cite{Reuter}, where this case is taken
as a base for a bundle approach to quantum mechanics. Our generic opinion is
that the phase space is not a `suitable' candidate for a bundle's base, the
reason being the Heisenberg uncertainty principle by virtue of which the
points of the phase space have no physical
meaning~\cite[chapter~IV]{Messiah-1}. This reason does not apply if as a base
is taken the phase space of some observer as, by definition, the observers
are treated as classical objects (systems). Therefore one can set the base
$\base$ of the Hilbert bundle $\HilB$ to be the phase space of some observer.
Then the reference path $\gamma\colon J\to\base$ can be interpreted as the
observer's phase\ndash space trajectory which, generally, can have self\ndash
intersections. The further treatment of this case is the same as of
$\base=\mathbb{E}^3$.
	Regardless of the above\ndash said, one can always identify $\base$
with the system's phase space, if it exists, as actually $\base$ is a free
parameter in the present work.

	An interesting bundle approach to quantum mechanics
is contained in~\cite{Reuter}. In it the evolution of a quantum system is
described in a Hilbert bundle \emph{over the system's phase space} with the
ordinary system's Hilbert space as a (typical) fibre which is, some times,
identified with the fibre over an arbitrary fixed phase-space point. The
evolution itself is presented as a parallel transport in the bundle space
generated via non\ndash dynamical linear (and symplectic) connection which is
closely related to the symplectic structure of the phase space.
	An important feature of~\cite{Reuter} is that in it the bundle
structure is derived from the physical content of the paper. In this
sense~\cite{Reuter} can be considered as a good motivation for the general
constructions in this investigation.

	Before comparing the mathematical results of~\cite{Reuter} with the
ones of this work, we have to say that the \emph{loc.\ cit.}\ contains some
incorrect `bundle' expressions which, however, happily do not influence most
of the conclusions made on their base.
       In~\cite{bp-Reuter-comments} we point to and show possible ways for
improving of a number of mathematically non-rigorous or wrong expressions,
assertions, and definitions in~\cite{Reuter}. We emphasize that all this
concerns only the `bundle' part of the mathematical structure of \emph{loc.\
cit.}\ and does not deal with its physical contents.
	The general moral of the critical remarks
of~\cite{bp-Reuter-comments} is: most of the final results and conclusions
of~\cite{Reuter} are valid provided the pointed in~\cite{bp-Reuter-comments}
(and other minor) corrections are made. Below we shall suppose that this
is carefully done. On this base we will compare~\cite{Reuter} with the
present work.

	The main common point between~\cite{Reuter} and this investigation is
the consistent application of the fibre bundle theory to (nonrelativistic)
quantum mechanics. But the implementation of this intention is quite
different: in~\cite{Reuter} we see a description of quantum mechanics in a
new, but `frozen', geometrical background based on a non-dynamical linear
connection deduced from the symplectical structure of the system's phase
space, while the present work uses a `dynamical geometry' (linear transport
along paths, which may turn to be a parallel one generated by a linear
connection) whose properties depend on the system's Hamiltonian, \ie on the
physical system under consideration itself.

	The fact that in~\cite{Reuter} the system's phase space is taken as a
base of the used Hilbert bundle is not essential since nothing can prevent us
from making the same choice as, actually, the base is not fixed here.
In~\cite{Reuter}  is partially considered the dynamics of multispinor fields.
This is an interesting problem, but, since it is not primary related to
conventional quantum mechanics, we think it is out of the scope of our work.
The methods of its solution are outlined in~\cite{Reuter} and can easily be
incorporated within our bundle quantum mechanics.

	The fields of (metaplectic) spinors used in~\cite{Reuter} are simply
sections of the Hilbert bundle, while the ``world-line spinors'' in
\emph{loc.\ cit.}\ are sections along paths in our terminology. The family of
operators $\mathcal{O}_{\phi^a}$ or
$\mathcal{O}_{f}(\phi)$~\cite[equations~(4.8) and~(4.9)]{Reuter} acting on
$\bHil_\phi$ are actually bundle morphisms.

	A central r\^ole in both works plays the `principle of invariance of
the mean values': the mean values (mathematical expectations) of the
morphisms corresponding to the observables  (dynamical variables) are
independent of the way they are calculate. We have used this assumption many
times (see, e.g., Sections~\ref{VI}, \ref{VII}, and~\ref{XI}, in particular,
equations~\eref{6.1''}, \eref{7.5}, \eref{7.11}, \eref{7.17},
and~\eref{11.17}) without explicitly formulating it as a `principle'. But if
one wants to build axiomatically the bundle quantum mechanics, he will be
forced to include this principle (or an equivalent to it assertion) into the
basic scheme of the theory. In~\cite{Reuter}  `the invariance of the mean
values' is mentioned several times and it is used practically in the form of
the `background-quantum split symmetry' principle, explained
in~\cite[sect.~4]{Reuter}
(see, e.g.,~\cite[equation~(4.18)]{Reuter} and the comments after it).
Its particular realizations are written
as~\cite[equation~(4.17) and second equation~(4.42)]{Reuter}
which are equivalent to it in the corresponding context. A consequence of the
mean-value invariance is the `Abelian' character of the compatible with it
connections, expressed by~\cite[equation~(4.14)]{Reuter}, which is a special
case of our result~\cite[equation~(4.4)]{bp-BQM-preliminary}.
In~\cite{Reuter} the mean values are independent of the point at which they
are determined. In our bundle quantum mechanics this is not generally the case
as different points correspond to different time values (see,
e.g.,~\eref{6.1''}). This difference clearly reflects the dynamical character
of our approach and the `frozen' geometrical one of~\cite{Reuter}. In any
case, the principle we are talking about is so important that without it
the equivalence between the bundle and conventional forms of quantum
mechanics can not be established.

	In both works the quantum evolution is described via appropriate
transport along paths:
In~\cite[see, e.g., equations~(3.54) and~(4.53)]{Reuter} this is an `Abelian'
parallel transport along curves, whose holonomy group is
$U(1)$~\cite[equation~(4.38)]{Reuter}, while in our investigation is employed
a transport along paths uniquely determined by the Hamiltonian
(see Sect.~\ref{IV}) which, generally, need not to be a parallel translation.

	Now we turn our attention on the bundle equations of motion: in the
current work we have a \emph{single} bundle Schr\"odinger
equation~\eref{5.13} (see also its matrix version~\eref{5.1}), while
in~\cite[equation~(5.54)]{Reuter} there is an \emph{infinite} number of such
equations, one Schr\"odinger equation in each fibre $\bHil_\phi$ for the
system's state vector $|\psi(t)\rangle_\phi$ at every point $\phi\in\base$.%
\footnote{%
Note that the appearing in~\cite[equations~(4.54)--(4.56)]{Reuter} operator
$\mathcal{O}_H$ is an analogue of our matrix-bundle Hamiltonian
(see Sect.~\ref{V}).%
}
Analogous is the situation with the statistical operator (compare our
equation~\eref{11.17} or~\eref{11.15} with~\cite[equation~(4.56)]{Reuter}).
This drastical difference is due to the \emph{different objects} used to
describe systems states: for the purpose we have used \emph{sections along
paths} (see Sect.~\ref{new-I}), while in~\cite{Reuter} are utilized
\emph{(global) sections} of the bundle defined via~\eref{4.3b}
(cf.~\cite[equation~(4.41)]{Reuter}). Hence, what actually is done
in~\cite{Reuter} is the construction of an isomorphic images of the quantum
mechanics from the fibre $\Hil$ in every fibre $\bHil_\phi$, $\phi\in\base$
(see the comments after~\eref{4.3b}).

	To summarize the comments on part of the mathematical
structures in~\cite{Reuter}: It contains a fibre
bundle description of quantum mechanics. The state vectors are replaced by
(global) sections of a Hilbert bundle with the system's phase space as a base
and their (bundle) evolution is governed through Abelian parallel transport
arising from the symplectical structure of the phase space. Locally, in any
fibre of the bundle, the evolution is presented by a Schr\"odinger
equation, specific for each fibre of the bundle. The work contains a number
of incorrect mathematical constructions which, however, can be corrected so
that the final conclusions remain valid. Some ideas of the paper are near to
the ones of this investigation but their implementation and development is
quite different in both works.

\section{Summary}
\label{summary}
\setcounter{equation} {0}

	In this work we have proposed and developed  a new invariant
fibre bundle formulation of nonrelativistic quantum mechanics. It is fully
equivalent to the usual formulation of the theory in the case when the
underlying manifold is the (coordinate) three dimensional Euclidean space of
classical (quantum) mechanics. In the new description a pure state of a
quantum system is describe by a state section along paths of a Hilbert fibre
bundle.  The time evolution of the state sections obeys the bundle
Schr\"odinger equation~(\ref{5.13}). A mixed state of a quantum system is
described via the density morphism along paths satisfying the Schr\"odinger
(type) equation~\eqref{11.17}. In the proposed bundle approach to any
dynamical variable corresponds a unique observable which is a bundle morphism
along paths of the Hilbert fibre bundle of the investigated system.  The
observed value of a dynamical variable is equal (by definition) to the mean
value (the mathematical expectation) of the corresponding bundle morphism and
it is calculated by means of the bundle state section or density morphism
corresponding to the system's state at the moment.

	The correspondence between the conventional Hilbert space description
and the new Hilbert bundle description of non-relativistic quantum
mechanics is given in table~\ref{table:space/bundle} on
page~\pageref{table:space/bundle}.

	A feature of the bundle form of quantum mechanics is the inherent
connection between physics and geometry: the system's Hamiltonian
(via equation~\eref{5.10}) completely determines the concrete properties of
the system's Hilbert bundle. In this sense, here the Hamiltonian plays the
same r\^ole as the energy\ndash momentum tensor in general relativity.
Another view\nobreakdash-point (also based on~\eref{5.10}) is to look on the
Hamiltonian as a gauge field in the sense of Yang-Mills theories. In any
case, we see in the bundle quantum mechanics a realization of the
intriguing idea, going back to Albert Einstein and Bernhard Riemann,
that the physical properties of the systems are responsible
for the geometry of the spaces used for their description.

\section{Discussion}
\label{discussion}
\setcounter{equation} {0}

	Since the set of all sections of a vector bundle is a module over the
ring of all ($C^0$) functions on its base with values in the field with
respect to which it has a vector
structure~\cite[chapter~3, propositions~1.6]{Husemoller},
the set $\Sec{\HilB}$ is a module over the ring of functions
$f\colon\base\to\mathbb{C}$.
Besides, this module is equipped with a scalar product. In
fact, if $\Phi,\Psi\in\Sec{\HilB}$, and $f,g\colon\base\to\mathbb{C}$, the
vector structure of $\Sec{\HilB}$ is given by
	\begin{equation}	\label{concl.1}
(f\Phi+g\Psi)\colon x\mapsto f(x)\Phi(x) + g(x)\Psi(x) \in\bHil_x,
\qquad x\in\base
	\end{equation}
and its inner product is defined via
 $(\Phi,\Psi)\mapsto\langle\Phi | \Psi\rangle\colon \base \to \mathbb{C}$
where
	\begin{equation}	\label{concl.2}
\langle\Phi | \Psi\rangle \colon x\mapsto \langle\Phi | \Psi\rangle (x) :=
\langle\Phi(x) | \Psi(x)\rangle_x \in\mathbb{C}, \qquad x\in\base.
	\end{equation}
Such a structure, module with an inner product, can naturally be called a
\emph{Hilbert module}.

	Moreover, any bundle morphism $\mor{A}\in\Morf{\HilB}$ can be
considered as an operator $\mor{A}\colon\Sec{\HilB}\to\Sec{\HilB}$ of the
sections over $\HilB$ whose action is defined by
	\begin{equation}	\label{concl.3}
(\mor{A}\Phi)\colon x\mapsto\mor{A}_x(\Phi(x)),\qquad
\Phi\in\Sec{\HilB},\quad x\in\base
	\end{equation}
and vice versa, to any operator $\mor{B}\colon\Sec{\HilB}\to\Sec{\HilB}$
there corresponds a unique morphism of $\HilB$ whose restriction $\mor{B}_x$
on $\bHil_x$ is given via
	\begin{equation}	\label{concl.4}
\mor{B}_x(\Phi(x)):=(\mor{B}\Phi)(x),\qquad
\Phi\in\Sec{\HilB},\quad x\in\base.
	\end{equation}

	These mathematical results allow us, if needed, to reformulate
(equivalently) the Hilbert bundle description of quantum mechanics in terms
of vectors and operators, but now in the Hilbert module of sections of the
Hilbert fiber bundle over the space-time.

	Since any pure state of a quantum system can be described via
a suitable density operator~\cite[chapter~VIII, \S~24]{Messiah-QM}, the remark
at the end of subsection~\ref{XI.2} suggests that it is possible the
nonrelativistic quantum mechanics to be formulated entirely in terms of
morphisms along paths in the fibre bundle of the morphisms along paths of the
Hilbert bundle of the considered system.

	In the present investigation we have not fixed the base $\base$ of the
Hilbert bundle $\HilB$. We did not used even some concrete
assumptions about $\base$, except the self-understanding ones, e.g. such as
that it is an non-empty topological space. At the beginning of
Sect.~\ref{new-I} we assumed $\base$ to be the (coordinate) 3-dimensional
Euclidean space $\mathbb{E}^3$ of quantum (or classical) mechanics. This is
required for the physical interpretation of the developed here theory. This
interpretation holds true also for any differentiable manifold $\base$ with
$\dim \base \ge 3$. This is important in connection with further
generalizations. For instance, such are the cases when $\base$ is chosen as
the 4-dimensional Minkowski space-time $M^4$ of special relativity, or
the pseudo-Riemannian space-time $V_4$ of general relativity, or the
Riemann-Cartan space-time $U_4$ of the  $U_4$-gravitational theory.
All this points to the great arbitrariness in the choice of the
geometrical structure of~$\base$. Generally it has to be determined by a
theory different from quantum mechanics, such as the classical
mechanics  or the special or general relativity. As a consequence of
this, there is a room for some kind of unified theories, for instance
for a unification of quantum mechanics and general relativity.
These problems will be investigated elsewhere.

	The developed in the present investigation theory is global in a sense
that in it we interpret $\base$ as being the whole space(-time) where the
studied objects `live'. If by some reasons one wants or is forced to consider
a system resided into a limited region of $\base$, then all the theory can be
localized by replacing $\base$ with this region or simply via
\emph{mutatis mutandis} restricting the already obtained results on it. At
the same time, our theory is local in a sense that such are the used in it
observers which are assumed to can make measurements only at the points they
reside. The theory can slightly be generalized by admitting the observers can
perform observations (measurements) at points different from their own
residence. This puts the problem of defining the mean values of the
observables at points different from the one at which the observer is
situated in such a way as they to be independent of the possibly introduce for
this purpose additional constrictions
(cf.~\cite[sect.~3]{bp-BQM-preliminary}).  This problem will be solved
elsewhere.

	The observers we have been dealing until now in this work can be
called scalar and point-like as it is supposed that they have no internal
structure and the only their characteristics in the theory are their positions
(generally) in the space-time $\base$. Notice that as a set the manifold
$\base$ coincides with the variety of all possible positions of all possible
observers. This observation suggests that in the general case $\base$ has to
be replaced with a set~ $\widetilde\base$ consis-
\newpage%
\thispagestyle{plain}%
\addtolength{\oddsidemargin}{-6em}
\include{bqm-tab}
\newpage%
\addtolength{\oddsidemargin}{+6em}
\noindent%
ting of all of the values of
all parameters describing completely the states of any possible observer.
Naturally, the set $\widetilde\base$ has to be endowed with some topological
or/and smooth geometrical structure. For instance, consider an `anisotropic'
point-like observers characterized by their position $x$ in $\base$ and some
vector $\mor{V}_x$ at it, i.e.\  by pairs like $(x,\mor{V}_x)$, $x\in\base$,
$\mor{V}_x\in \mor{L}_x$ with $\mor{L}_x$ being some vector space. In this
case $\widetilde\base$ is naturally identified with the (total) bundle space
$\mor{L}$ of the arising over $\base$ vector bundle
$(\mor{L},\pi_\mor{L},\base)$ with
$\mor{L}:=\bigcup_{x\in\base}\bigcup_{\mor{V}_x\in\mor{L}_x}(x,\mor{V}_x)$
and
$\pi_\mor{L}(x,\mor{V}_x):=x\in\base$,
i.e.\  we can put $\widetilde\base=\mor{L}$. In particular, if $\mor{V}_x$
is the observer's velocity at $x$, we have $\widetilde\base=T(\base)$
with $(T(\base),\pi_{T(\base)},\base)$
being the bundle tangent to $\base$. Another sensible example is when
$\mor{V}_x$ is interpreted as observer's spin etc.
	In connection with some recent investigations (see,
e.g.,~\cite{Ashtekar&Schilling,Reuter}) it is worth to be studied the special
case when $\widetilde\base$ is taken to be the (classical) phase space of the
observer.
Returning to the general
situation, we see that our theory can easily be modified to cover such
generalizations. For this purposed we have simply to replace the Hilbert
bundle $\HilB$ over $\base$ with the Hilbert bundle
$(\bHil,\proj,\widetilde\base)$ over $\widetilde\base$.
This, together with other evident corresponding changes, such as
$(x\in\base)\mapsto (x\in\widetilde\base)$ and
 $(\gamma\colon J\to\base)\mapsto(\gamma\colon J\to\widetilde\base)$,
allows us to apply the developed in the present investigation Hilbert bundle
description to far more general situations than the one we were speaking
about until now. In principle, the afore-described procedure is applicable to
non-local observers too, but this is out of the subject of the present work.

	We want also to mention that since any fibre of the Hilbert bundle
$\HilB$ is an isomorphic image of the Hilbert space $\Hil$, the conventional
probabilistic interpretation of the nonrelativistic quantum
mechanics~\cite{Messiah-QM,Fock-FQM} remains
\emph{mutatis mutandis} completely
valid in the Hilbert bundle description too. For this purpose one has to
replace state vectors and operators acting on $\Hil$ with the corresponding
state sections along paths and morphisms along paths of $\HilB$ and then to
follow the general rules outlined in this work and, for instance,
in~\cite{Messiah-QM}.

	In connection with further applications of the bundle approach to the
quantum field theory, we notice the following. Since in this theory
the matter fields are represented by operators acting on (wave) functions
from some space, the matter fields in their bundle modification should be
described via morphisms (along paths) of a suitable fibre bundle whose
sections (along paths) will represent the wave functions. We can also,
equivalently, say that in this way the matter fields would be sections of the
fibre bundle of bundle morphisms of the mentioned suitable bundle. An
important point here is that the matter fields are primary related to the
bundle arising over the space-time (or other space which includes it) and
not to the space-time itself to which other structures are directly related,
such as connections and the principle bundle over it.


	The bundle approach to nonrelativistic quantum mechanics, developed in
the present investigation, seems also applicable to classical mechanics,
statistical mechanics, relativistic quantum mechanics, and field theory. We
hope that such a novel treatment of these theories will reveal new
perspectives for different generalizations, in particular for their
unification with the theory of gravitation.

\addcontentsline{toc}{section}{References}

\bibliography{bozhopub,bozhoref}
\bibliographystyle{unsrt}

\addcontentsline{toc}{subsubsection}{\vspace{1ex}This article ends at page}

\end{document}

%% file: bqm-tab.tex
\renewcommand{\arraystretch}{1.2}
%
%
%
\begin{table}
\vspace{-20.0ex}
\footnotesize
\centering
\caption{\normalsize
Comparison between Hilbert space and Hilbert bundle descriptions.
\vspace{2.2ex} }
\label{table:space/bundle}
\begin{tabular}{||l|l|l||}
\hline
\hline
\multicolumn{1}{||c||}
{
\begin{tabular}{c}
\small 
\\ \bfseries Hilbert space description \\ \\
\end{tabular}
} &
\multicolumn{1}{||c||}
{
\begin{tabular}{c}
\small 
\\ \bfseries Hilbert bundle description \\ \\
\end{tabular}
} &
\multicolumn{1}{||c||}
{
\begin{tabular}{c}
\small 
\\ \bfseries Remark(s) \\ \\
\end{tabular}
}
\\  
 \hline \hline
Hilbert space $\Hil$ &
Hilbert fibre bundle $\HilB$ &
$l_x:\bHil_x\to\Hil$
\\ \cline{1-3}
Vector $\varphi\in\Hil$ &
Section $\overline{\Phi}\in\Sec{\HilB}$ &
$\overline{\Phi}\colon x \mapsto l_x^{-1}(\varphi)$
\\ \cline{1-3}
Operator $\ope{A}:\Hil\to\Hil$ &
Bundle morphism $\overline{\mor{A}}\in\Morf{\HilB}$ &
$\overline{\mor{A}}_x=l_x^{-1}\circ\ope{A}\circ l_x$
\\ \cline{1-3}
State vector $\psi\in\Hil $ &
State section $\Psi$ along paths of $\HilB$ &
$\Psi_\gamma (t)=l_{\gamma(t)}^{-1}(\psi(t))$
\\ \cline{1-3}
Observable $\ope{A}\colon\Hil\to\Hil$ &
Bundle morphism $\mor{A}$ along paths of $\HilB$ &
$\mor{A}_\gamma(t)=l_{\gamma(t)}^{-1}\circ\ope{A}(t)\circ l_{\gamma(t)}^{}$
\\ \cline{1-3}
Hermitian scalar product $\langle\phi | \psi\rangle$
&
Hermitian bundle scalar product
  $\langle\Phi_x | \Psi_x\rangle_x$
&
$\langle\cdot | \cdot\rangle_x = \langle l_x\cdot | l_x \cdot\rangle$
\\ \cline{1-3}
\multicolumn{1}{||l|}
{
\begin{tabular}{@{}l@{}}
Hermitian conjugate operator     \\
  $\ope{A}^\dag$ to an operator $\ope{A}$:         \\
     $\langle\ope{A}^\dag\phi | \psi\rangle =
   \langle \phi | \ope{A}\psi\rangle$
\end{tabular}
}
&
\multicolumn{1}{l|}
{
\begin{tabular}{@{}l@{}}
1) Hermitian conjugate map
   $\mor{A}^\ddag_{x\to y}\colon\bHil_x\to\bHil_y$    \\
\ \ \ \       to a bundle map $\mor{A}_{y\to x}:\bHil_y\to \bHil_x$: \\
\ \ \ \   $\langle \mor{A}_{x\to y}^\ddag\Phi_x | \Psi_y\rangle_y =
 \langle\Phi_x | \mor{A}_{y\to x} \Psi_y\rangle_x$ \\[1 ex]
2)  Hermitian conjugate morphism $\mor{A}^\ddag$   \\
\ \ \ \    to a bundle morphism $\mor{A}$ along paths: \\
\ \ \ \   $\langle \mor{A}_x^\ddag\Phi_x | \Psi_x\rangle_x =
  \langle\Phi_x | \mor{A}_x \Psi_x\rangle_x$
\end{tabular}
}
&
\multicolumn{1}{l||}
{
\begin{tabular}{@{}l@{}}
$\mor{A}_{x\to y}^\ddag = $ \\
$ = l_y^{-1}\circ(l_x\circ \mor{A}_{y\to x}\circ l_y^{-1})^\dag \circ l_x$
\\
\\[1 ex]
\\
$\mor{A}_x^\ddag = l_x^{-1}\circ(l_x\circ \mor{A}_x\circ l_x^{-1})^\dag \circ l_x$
\end{tabular}
}
\\ \cline{1-3}
Unitary operator: $\ope{A}^\dag=\ope{A}^{-1}$ &
\multicolumn{1}{|l|}
{
\begin{tabular}{@{}l@{}}
1) Unitary  bundle map:  \\
\ \ \ \   $\mor{A}_{x\to y}^\ddag=\mor{A}_{y\to x}^{-1}:=
    (\mor{A}_{y\to x})^{-1}$
\\[1ex]
2) Unitary bundle morphism: $\mor{A}^\ddag=\mor{A}^{-1}$
\end{tabular}
}
&
 \multicolumn{1}{l||}
{
\begin{tabular}{@{}l@{}}
$\mor{A}_{x\to y}^{-1}$ is the left   \\
inverse of $\mor{A}_{x\to y}$  \\[1ex]
$\mor{A}_x^\ddag=\mor{A}_x^{-1}$
\end{tabular}
}
\\ \cline{1-3}
Hermitian operator: $\ope{A}^\dag=\ope{A}$ &
Hermitian morphism: $\mor{A}^\ddag=\mor{A}$ &
$\mor{A}_x^\dag=\mor{A}_x$;
$\mor{A}^\ddag=\mor{A}\iff\ope{A}^\dag=\ope{A}$
\\ \cline{1-3}
Basis $\{f_a(t)\}$ in $\Hil$ &
Field of bases $\{e_a\}$ in $\Sec\HilB$ &
\begin{tabular}{@{}l@{}}
A good choice is \\ $e_a(x)=l_x^{-1}(f_a(t))$
\end{tabular}
\\ \cline{1-3}
\multicolumn{2}{||l|}
{
\begin{tabular}{@{}l@{}}
Matrix corresponding
to a linear map or a vector in a given basis (bases):
\\
the same notation but the kernel letter is in \textbf{boldface}
\end{tabular}
}
&
\multicolumn{1}{l||}
{
\begin{tabular}{@{}l@{}}
For example:
\(
\boldsymbol{\ope{A}}(t),\ \mmor{A},\
\boldsymbol{\psi}(t),\)
\\
\(
\boldsymbol{\Psi}_\gamma(t),\
\boldsymbol{l}_x,\
\mope{U}(t,s),\  \mmor{U}_\gamma(t,s)
\)
\end{tabular}
}
\\ \cline{1-3}
\begin{tabular}{@{}l@{}}
Mean value of an operator $\ope{A}$: \\
  $\langle \ope{A}(t)\rangle_\psi^t =
  \frac{\langle \psi(t) | \ope{A}(t)\psi(t)\rangle}
  {\langle \psi(t) | \psi(t)\rangle}$
\end{tabular}
&
\begin{tabular}{@{}l@{}}
Mean value of a bundle morphism $\mor{A}(t)$: \\
\(
\langle \mor{A}(t)\rangle_{\Psi_\gamma}^t =
\frac{\langle \Psi_\gamma(t) | \mor{A}(t)\Psi_\gamma(t)\rangle_{\gamma(t)}}
{\langle \Psi_\gamma(t) | \Psi_\gamma(t)\rangle_{\gamma(t)} }
\)
\end{tabular}
&
$\langle \mor{A}(t)\rangle_{\Psi_\gamma}^t =
   \langle \ope{A}(t)\rangle_\psi^t$
\\ \cline{1-3}
Evolution operator $\ope{U}$ &
Evolution transport $\mor{U}$ along paths &
 \multicolumn{1}{l||}
{
\begin{tabular}{@{}l@{}}
$\mor{U}_\gamma(t,s)=l_{\gamma(t)}^{-1}\circ\ope{U}(t,s)\circ l_{\gamma(s)}$;
\\
see~\eqref{2.1} and~\eqref{4.4}
\end{tabular}
}
\\ \cline{1-3}
1) Hamiltonian ${\Ham}$ &
1) Bundle Hamiltonian $\bHam$ &
$\bHam_\gamma(t)=l_{\gamma(t)}^{-1}\circ\Ham(t)\circ l_{\gamma(t)}$ \\
2) Matrix Hamiltonian $\mHam$ &
2) Matrix-bundle Hamiltonian $\mbHam_\gamma$ &
See~(\ref{5.08}), (\ref{5.2}), and~(\ref{5.2'}) \\[1ex]
 &
\hspace*{-0.65em}\begin{tabular}{l}
3) Matrix $\boldsymbol{\Gamma}$ of the coefficients  \\
\ \ \ \ of the evolution transport
\end{tabular}
&
$\boldsymbol{\Gamma}_\gamma(t)=-\mbHam_{\gamma}(t)/\ih$
\\ \cline{1-3}
\multicolumn{1}{||l|}
{
\begin{tabular}{@{}l@{}}
1) Schr\"odinger equation: \\
\ \ \ \  \( \ih \frac{d \psi(t)}{d t} = \Ham(t)\psi(t) \) \\[1ex]
2) Matrix Schr\"odinger equation: \\
\ \ \ \   \( \ih \frac{d \boldsymbol{\psi}(t)}{d t} =
  \mHam(t)\boldsymbol{\psi}(t) \)
\end{tabular}
}
&
\multicolumn{1}{l|}
{
\begin{tabular}{@{}l@{}}
1) Bundle Schr\"odinger equation: \\
\ \ \ \  $\mor{D}_t^\gamma \Psi_\gamma = 0$    \\[1ex]
2) Matrix-bundle Schr\"odinger equation: \\
\ \ \ \    \( \ih \frac{d \boldsymbol{\Psi}_\gamma(t)}{d t} =
 \mbHam_{\gamma}(t)\boldsymbol{\Psi}_\gamma(t) \)
\end{tabular}
}
&
\multicolumn{1}{l||}{Equivalent equations}
\\ \cline{1-3}
Density operator $\rho$ &
Density morphism $\Rho$ along paths &
 $\Rho_\gamma(t)=l_{\gamma(t)}^{-1}\circ\rho(t)\circ l_{\gamma(t)}$
\\ \cline{1-3}
\begin{tabular}{@{}l@{}}
Density operator evolution: \\
$ \ih\frac{d\rho(t)}{d t} = [\Ham(t),\rho(t)]_{\_} $
\end{tabular}
&
\begin{tabular}{@{}l@{}}
Density morphism evolution: \\
$ \widetilde{\mor{D}}_{t}^{\gamma}(\widetilde{\Rho}_\gamma) = 0 $ or
$ ^\circ\!{\mor{D}}_{t}^{\gamma}(\Rho_\gamma) = 0 $
\end{tabular}
&
Equivalent equations
\\ \cline{1-3}
{
\begin{tabular}{@{}l@{}}
Schr\"odinger picture of motion: \\
$\psi(t),\ \ope{A}(t)$; \,  $\rho(t),\ \ope{A}(t)$
\end{tabular}
}
&
\multicolumn{1}{l|}
{
\begin{tabular}{@{}l@{}}
Bundle Schr\"odinger picture of motion: \\
$\Psi_\gamma(t),\ \mor{A}_\gamma(t)$; \,
$\Rho_\gamma(t),\ \mor{A}_\gamma(t)$
\end{tabular}
}
&
\begin{tabular}{@{}l@{}}
See:~(\ref{2.5}), (\ref{5.13}); \\
  \eqref{11.5}, \eqref{11.15}, \eqref{11.17}
\end{tabular}
\\ \cline{1-3}
\multicolumn{1}{||l|}
{
\begin{tabular}{@{}l@{}}
Heisenberg picture of motion:  \\
$\psi_t^{\mathrm{H}}(t_0),\ \ope{A}_t^{\mathrm{H}}(t_0)$; \,
$\rho_t^{\mathrm{H}}(t_0),\ \ope{A}_t^{\mathrm{H}}(t_0)$
\end{tabular}
}
&
\multicolumn{1}{l|}
{
\begin{tabular}{@{}l@{}}
Bundle Heisenberg picture of motion: \\
$\Psi_{\gamma,t}^{\mathrm{H}}(t_0),\ \mor{A}_{\gamma,t}^{\mathrm{H}}(t_0)$; \,
$\Rho_{\gamma,t}^{\mathrm{H}}(t_0),\ \mor{A}_{\gamma,t}^{\mathrm{H}}(t_0)$
\end{tabular}
}
&
\begin{tabular}{@{}l@{}}
See:~(\ref{7.11a}), (\ref{7.8}), (\ref{7.13}),\\ (\ref{7.13a});
\eqref{11.18},~\eqref{11.19}
\end{tabular}
\\ \cline{1-3}
\multicolumn{1}{||l|}
{
\begin{tabular}{@{}l@{}}
`General' picture of motion:  \\
$\psi_t^\ope{V}(t_0),\ \ope{A}_t^\ope{V}(t_0)$; \,
$\rho_t^\ope{V}(t_0),\ \ope{A}_t^\ope{V}(t_0)$
\end{tabular}
}
&
\multicolumn{1}{l|}
{
\begin{tabular}{@{}l@{}}
Bundle `general' picture of motion:  \\
$\Psi_{\gamma,t}^\mor{V}(t_0),\ \mor{A}_{\gamma,t}^\mor{V}(t_0)$; \,
$\Rho_{\gamma,t}^\mor{V}(t_0),\ \mor{A}_{\gamma,t}^\mor{V}(t_0)$
\end{tabular}
}
&
\begin{tabular}{@{}l@{}}
See:~(\ref{7.24}), (\ref{7.27})--(\ref{7.29}),\\ (\ref{7.20}); 
\,~\eqref{11.26}--\eqref{11.29}
\end{tabular}
\\ \cline{1-3}
\begin{tabular}{@{}l@{}}
Integral of motion $\ope{A}$:\\
$\ih \frac{\partial \ope{A}(t)}{\partial t} +[\ope{A}(t),\Ham(t)]_{\_} = 0$
\end{tabular}
&
\begin{tabular}{@{}l@{}}
Integral of motion $\mor{A}$: \\
$\widetilde{\mor{D}}_{t}^{\gamma}\left(\mor{A}_\gamma\right)=0$ or
$^\circ\!{\mor{D}}_{t}^{\gamma}\left(\mor{A}_\gamma\right)=0$
\end{tabular}
&
\begin{tabular}{@{}l@{}}
Equivalent concepts. See:~\eqref{8.9},\\
\eqref{8.17};\,~\eqref{8.10},~\eqref{8.16}
\end{tabular}
\\ \hline\hline
\end{tabular}
\end{table}
%
%